# Functional Dynamics of PDZ Binding Domains:
# A Normal Mode Analysis

[1*]Paolo De Los Rios, [2]Fabio Cecconi, [3]Anna Pretre, [3]Giovanni Dietler, [4]Olivier Michielin, [1]Francesco Piazza, and [1]Brice Juanico.

[1] Laboratoire de Biophysique Statistique, ITP, Ecole Polytechnique Fédérale de Lausanne, 1015 Lausanne, Switzerland.

[2] INFM Università di Roma « La Sapienza », P.le Aldo Moro 2, 00185 Roma.

[3] Laboratoire de Physique de la Matière Vivante, IPMC, Ecole Polytechnique Fédérale de Lausanne, 1015 Lausanne, Switzerland.

[4] Institut Suisse de Recherche Expérimentale sur le Cancer (ISREC), Ch. des Boveresses 155, 1066 Epalinges/Lausanne, Switzerland.

*Corresponding author : Paolo.DeLosRios@epfl.ch

**PDZ (Post-synaptic density-95/discs large/zonula occludens-1) domains are relatively small (80 to 120 residues) protein binding modules central in the organization of receptor clusters and in the association of cellular proteins. Their main function is to bind C-terminals of selected proteins that are recognized through specific amino-acids in their carboxyl end. Binding is associated with a deformation of the PDZ native structure and is responsible for dynamical changes in regions not in direct contact with the target. We investigate how this deformation is related to the harmonic dynamics of the PDZ structure and show that one low-frequency collective normal mode, characterized by the concerted movements of different secondary structures, is involved in the binding process. Our results suggest that even minimal structural changes are responsible of communication between distant regions of the protein, in agreement with recent Nuclear Magnetic Resonance (NMR) experiments. Thus PDZ domains are a very clear example of how collective normal modes are able to characterize the relation between function and dynamics of proteins, <u>and</u> to provide indications on the precursors of binding/unbonding events.**

## Introduction

PDZ interaction domains are fundamental in regulating the dynamic organization of the cell. They play a central role in signaling pathways by organizing networks of receptors and in targeting selected cellular proteins to multiprotein complexes (Hung and Sheng 2002; Pawson and Nash 2003; Sheng and Sala 2001; Zhang and Wang 2003): for example the scaffolding protein InaD (Inactivation no afterpotential D) contains five PDZ domains that keep multiple proteins of the *Drosophila* photo-transduction cascade together (Tsunoda and Zuker 1999); the postsynaptic density protein PSD-95 (Post-synaptic density-95) comprises three PDZ domains that cluster glutamate receptors at synapses (Hung and Sheng 2002; Sheng and Sala 2001; Zhang and Wang 2003). In this way, the concerted action of these multiple proteins is made more efficient. The PDZ domain is also a fundamental component of the ZASP (Z-band alternatively spliced PDZ-motif) protein, required for maintaining the cytoskeletal structure during muscle contraction: the ZASP PDZ domain is responsible of the interaction between ZASP and the C-terminus of α-actinin-2 (Faulkner and others 1999). Overall, several hundreds of PDZ containing proteins have been identified in eukaryotes, from *S. cerevisiae* (yeast) to humans, and many of them contain multiple copies of the domain, making the PDZ one of the most abundant interaction domain yet identified.

Most PDZ-mediated interactions occur through the recognition of C-terminal peptide motifs (Hung and Sheng 2002; Sheng and Sala 2001), although growing evidence suggests that binding can be specific also for internal regions of the sequence mimicking the structure of a carboxyl terminal. This is the case of the nNOS (neuronal Nitric Oxide Synthase) and syntrophin complex, where a β-hairpin "finger" of nNOS docks in the PDZ syntrophin-binding groove (nNOS is itself a PDZ containing protein, so that this is a PDZ-PDZ complex)(Hillier and others 1999). Interestingly, the specificity of PDZ domains for different chemical and topological ligands relies on minor sequence variations, while the chemistry of the binding region and the overall domain fold are rather well conserved. Such high versatility of the fold, that with minor modifications allows targeting rather different partners, hints at some very favorable flexibility property of the three-dimensional structure of PDZ domains.

The dynamical characterization of PDZ domains is also beginning to emerge: very recently Fuentes *et al.* (Fuentes and others 2004) have argued, on the basis of Nuclear Magnetic Resonance (NMR) data, that the dynamics of PDZ domains upon ligand binding should show correlations over the entire protein structure.

Flexibility and internal mobility of proteins have been recognized as fundamental to their biological functions (Berendsen and Hayward 2000; Frauenfelder and others 2001; Kern and Zuiderweg 2003), since most of the biological activity occurs close to the native state. A powerful method to study their relation is Normal Mode Analysis (NMA). NMA had been originally proposed as a method alternative to Molecular Dynamics (MD) to explore the dynamics of proteins within their native basin (Go and others 1983; Levitt and others 1985; Noguti and Go 1982). NMA has seen a renaissance in very recent years, after the development of a few simplified methods that have proven to be very effective at describing the global, low frequency motions of proteins close to the native state (Atilgan and others 2001; Bahar and others 1997; Delarue and Sanejouand 2002; Doruker and others 2002; Hinsen 1998; Hinsen and others 1999; Marques and Sanejouand 1995; Sanejouand 1996; Suhre and Sanejouand 2004; Tama and others 2000; Tama and Sanejouand 2001; Tirion 1996). Clearly, whenever the protein function is related to some major allosteric transitions of the structure, Normal Modes, which do not explore conformations far from the native state, might not be appropriate to describe the relevant dynamics, and in such cases all-atom MD remains the best choice. However, in the PDZ case ligand binding does not induce large conformational changes of the domain structure and Normal Modes are expected to carry the relevant information for its function. We are furthermore confident that even in the case when larger structure deformations would be involved, Normal Modes may be the precursors for larger anharmonic motions.

In this work, we have mainly focused on the third PDZ domain of the PSD-95, hereafter referred to as PDZ3, whose structure has been determined by X-ray crystallography both with and without bound peptide (Doyle and others 1996). We also dealt with the PDZ domain of ZASP, recently solved by Nuclear Magnetic Resonance (NMR) using the X-ray structure of the PDZ3 of PSD-95 as a template (Au and others 2004). Finally we have analyzed the bound and unbound structures of the second PDZ domain of human protein tyrosine phosphatase 1E (hPTP1E), a soluble protein containing multiple PDZ domains that intervenes in a number of biological processes such as apoptosis, cytokinesis and signaling. Both structural and dynamic NMR data are available for hPTP1E (Kozlov and others 2002; Kozlov and others 2000).

## Methods

### Analysis of the X-ray and NMR data

We used the X-ray crystallographic structure of the third PDZ domain of PSD-95 both without and with bound peptide (1bfe and 1be9) from the Protein Data Bank (Berman and others 2000). The domain is 110 amino-acids long. Yet other PDZ domains range from 80 to 120 amino-acids, depending on where they have been excised from the sequence of the protein they belong to. Accordingly we have reduced the length of the PDZ3 cutting extremely mobile N- and C-terminal parts that may hinder our analysis by always showing the largest fluctuations amplitudes that are unrelated to the binding process. We have then truncated the structure from Arg-309 to Ser-393 (numbering according to 1bfe).

The vector of amino-acid displacements corresponding to the binding deformation has been obtained through standard Kabsch's alignment of the two structures 1bfe and 1be9 (Kabsch 1976). This amounts to finding the best linear transformation which brings one structure onto the other by minimizing the Root-Mean-Square Distance (RMSD) between the backbones of the complexed and peptide-free structures.

The NMR data for the ZASP PDZ domain structure (Au and others 2004) are available as 1rgw from the Protein Data Bank. The NMR data for the second PDZ domain of hPTP1E are available both in the complexed and peptide-free forms from the Protein Data Bank (3pdz and 1d5g respectively) and their difference has again been computed using Kabsch's algorithm. The eight amino-acids dynamically linked to the binding site, although not in direct contact with the bound peptide, have been listed by Fuentes *et al.* (Fuentes and others 2004).

### CHARMM Normal Modes Calculations

The VIBRAN module of the CHARMM program was used to determine the normal modes and normal mode frequencies by diagonalization of the force constant matrix. Normal modes were computed for fully minimized structures of the PDZ domain with a distance dependent dielectric, $\varepsilon=4r$, and a 16Å cutoff for the non-bonded interactions. The minimization was performed using the Steepest Descent algorithm until the root mean square of the energy gradient reached a value below $10^{-1}$ kcal/mol Å, followed by an Adopted Basis Newton-Raphson algorithm, until the gradient reached a value of $10^{-8}$ kcal/mol Å. This gradient value has been shown to be satisfactory for calculating normal modes and is expected to yield real frequencies modes (Tidor and Karplus 1994). No constraint was applied on the system.

The B-factors, proportional to the thermal fluctuation amplitudes of atoms around their equilibrium positions, were computed according to the standard formula

$$B_i = \frac{8\pi^2}{3}\left\langle \left|\Delta r_i\right|^2 \right\rangle = \frac{8\pi^2 k_B T}{3}\sum_n \frac{\left|\vec{u}_{n,i}\right|^2}{\omega_n^2} \tag{1}$$

where $k_B$ is Boltzmann's constant, $T$ the absolute temperature, $\omega_n$ the frequency of the $n$-th mode, and $\vec{u}_{n,i}$ the displacement of the $i$-th Cα atom in such a mode. The sum runs over non-zero frequency modes, thus discarding the six roto-translations.

## Cβ-ANM

The Anisotropic Network Model (ANM) (Atilgan and others 2001) is a coarse-grained model which instead of taking into account the full atomic details of the protein only considers its Cα carbons as representatives of the amino-acids (it is therefore a backbone-centric model). Chemically detailed interactions are simplified to harmonic interactions (springs) between Cα carbons closer than a given cutoff $R_c$. In the original formulation of the model the optimal $R_c$ value is around 13Å. However, the diameter of PDZ domains is not much larger than this cutoff and therefore the ANM would connect almost all amino-acids with each other. For this reason we preferred to use a recent version of the ANM, the Cβ Anisotropic Network Model (Cβ-ANM) (Micheletti and others 2004). In such a model, also Cβ carbons are considered (except for glycine whose side chain is simply a hydrogen) and the complex chemical interactions between residues are described by springs connecting all Cα-Cα, Cα-Cβ and Cβ-Cβ pairs whose distances, in the native fold, are smaller than $R_c=7.5$Å. The springs have stiffness $\gamma_{\alpha\alpha}$, $\gamma_{\alpha\beta}$ and $\gamma_{\beta\beta}$, irrespective of distance and chemical species of the residues. Thus the model is characterized by an energy function

$$E = \frac{\gamma_{\alpha\alpha}}{2}\sum_{i>j}\Delta\left(\left|\vec{R}_{i\alpha,j\alpha}\right|\right)\left(\left|\vec{r}_{i\alpha,j\alpha}\right|-\left|\vec{R}_{i\alpha,j\alpha}\right|\right)^2 + \gamma_{\alpha\beta}\sum_{i,j}\Delta\left(\left|\vec{R}_{i\alpha,j\beta}\right|\right)\left(\left|\vec{r}_{i\alpha,j\beta}\right|-\left|\vec{R}_{i\alpha,j\beta}\right|\right)^2$$
$$+ \frac{\gamma_{\beta\beta}}{2}\sum_{i>j}\Delta\left(\left|\vec{R}_{i\beta,j\beta}\right|\right)\left(\left|\vec{r}_{i\beta,j\beta}\right|-\left|\vec{R}_{i\beta,j\beta}\right|\right)^2 \tag{2}$$

where $\vec{R}_{i\alpha,j\alpha}$ indicates the vector distance between the $i$-th and $j$-th Cα atoms in the native structure and $\vec{r}_{i\alpha,j\alpha}$ is the same distance in a distorted conformation (and similarly for Cα-Cβ and Cβ-Cβ pairs). The function $\Delta\left(\left|\vec{R}_{i\alpha,j\alpha}\right|\right)$ is equal to one if the distance $\left|\vec{R}_{i\alpha,j\alpha}\right| < R_c$, zero otherwise, and it defines which Cα-Cα

pairs of atoms are in interaction, and analogously for Cα-Cβ and Cβ-Cβ pairs. The model eventually depends only on few parameters: $R_c$, determining the interaction range, and $\gamma_{\alpha\alpha}$, $\gamma_{\alpha\beta}$ and $\gamma_{\beta\beta}$, the interaction strengths, that are chosen so that $\gamma_{\alpha\alpha}=\gamma$ (and $\gamma_{\alpha\alpha}=2\gamma$ if the two Cα atoms are consecutive along the protein backbone) and $\gamma_{\alpha\beta}=\gamma_{\beta\beta}=\gamma/2$.

The implementation of the Cβ-ANM approach involves the diagonalization of the Hessian matrix obtained by the second derivatives of the energy function (3) with respect to Cα and Cβ displacements evaluated on the native structure coordinates (in complete analogy with the CHARMM approach). It is worth noticing that the native structure is the minimum of the energy by construction, so that no energy minimization stage is required. The Hessian matrix has 3Nx3N entries where N is the number of Cα's (85 in the truncated PDZ that we considered). The Cβ atoms coordinates are determined from the Cα coordinates with a small residual uncertainty (Micheletti and others 2004; Park and Levitt 1996), which is essentially irrelevant using such a simplified energy function as Equation 2. Therefore their degrees of freedom do not need to be explicitly taken into consideration, with the considerable advantage of not increasing the size of the Hessian matrix. The value of $\gamma$ is not derived from first principles, but it is adjusted through a least-square fitting of the computed B-factors to the experimental ones. The approximated vibrational spectrum of the protein is obtained from the eigenvalues of the Hessian matrix, and its eigenvectors are the Normal Modes.

**Thermal involvement coefficients**

A generic thermal fluctuation $\Delta\vec{r}$ of the *N*-residues structure around its crystallographic conformation can be expressed in a unique way as a superposition of normal modes,

$$\Delta\vec{r} = \sum_{n=1}^{3N-6} r_n \; \vec{u}_n \tag{3}$$

where $r_n$ is the contribution of the *n*-th normalized mode $\vec{u}_n$ to the fluctuation; the sum runs over all the *3N-6* non roto-translational modes, since we assume that $\Delta\vec{r}$ does not contain any roto-translational component.

The normalized vector $\vec{d}$ of the conformational deformation between complexed and uncomplexed structures of the truncated PDZ3 domain is defined as

$$\vec{d} = \frac{\vec{R}_{complex} - \vec{R}_{free}}{\left|\vec{R}_{complex} - \vec{R}_{free}\right|} \; . \tag{4}$$

The *3N* components of the vector $\vec{d} = \{d_{1x}, d_{1y}, d_{1z}, ..., d_{Nx}, d_{Ny}, d_{Nz}\}$ has *3N* components, represent each the displacements along the *x,y,z* directions of the Cα atoms. The normalization, $\left|\vec{d}\right|^2 = \sum_{i=1}^{N}\left(d_{ix}^2 + d_{iy}^2 + d_{iz}^2\right) = 1$, preserves the relative displacement amplitudes of different Cα atoms.

The overlap between a generic thermal fluctuation and the binding induced deformation is

$$\vec{d} \cdot \Delta\vec{r} = \sum_{n=1}^{3N-6} r_n \; \vec{d} \cdot \vec{u}_n = \sum_{n=1}^{3N-6} r_n \; I_n \tag{5}$$

where $I_n = \vec{d} \cdot \vec{u}_n$ is the so-called *involvement coefficient* (Ma and Karplus 1997; Tama and Sanejouand 2001; Xu and others 2003), that is the geometric projection of the deformation onto the *n*-th normal mode. Each mode-specific quantity $I_n$ determines how the *n*-th mode is geometrically similar to the deformation. The sum of the squares of the involvement coefficients is unitary by definition.

Modes with higher frequencies do not participate significantly to large-scale vibrational motions because their amplitudes decreases as $1/\omega$, a rule of thumb criterion used to focus on the low-frequency part of the vibrational spectrum of proteins in the search for functionally relevant modes (Ma and Karplus 1997). This physically correct but mathematically vague criterion can be cast in a rigorous formulation obtained from thermodynamic considerations. Since the average overlap $\left\langle \vec{d} \cdot \Delta\vec{r} \right\rangle$ vanishes because of symmetry, the physically relevant quantity to consider is the average of the square overlap,

$$\left\langle \left( \vec{d} \cdot \Delta\vec{r} \right)^2 \right\rangle = \sum_{n=1}^{3N-6} \left\langle r_n^2 \right\rangle I_n^2 = k_B T \sum_{n=1}^{3N-6} \left( \frac{I_n}{\omega_n} \right)^2 , \qquad (6)$$

where $\left\langle \bullet \right\rangle$ indicates the thermal average, which leads to the introduction of a new mode-specific indicator, the *thermal involvement coefficient* of mode *n*,

$$T_n = \frac{\left| I_n / \omega_n \right|}{\sqrt{\sum_{n=1}^{3N-6} \left( I_n / \omega_n \right)^2}} \qquad (7)$$

quantifying the contribution of the *n*-th normal mode to the overlap between the thermal fluctuations of the protein and the structural binding deformation $\vec{d}$ . Each $T_n$ depends on the inverse frequency of the corresponding mode *n* so that higher frequency modes intrinsically contribute less to the fluctuations. In this way thermodynamics automatically biases the search for functionally relevant modes to low-frequency ones. At the same time, Eq.7 dictates that that only modes with a sizeable involvement coefficient contribute appreciably to the function: this is the case for the PDZ normal modes in our analysis, where the lowest frequency NM is always almost irrelevant because of a small coefficient $I_n$ (Figures 3,5 and 9).

Weighting the involvement coefficients by their inverse frequency is therefore a non-subjective way to express the above mentioned common wisdom that high frequency modes do not contribute much to structure fluctuations.

## Results and Discussion

### *Crystallography of the binding deformations*

We have studied the truncated third PDZ domain (PDZ3) of PSD-95 (see Methods)(Doyle and others 1996; Hung and Sheng 2002; Sheng and Sala 2001), whose secondary structure consists of two α-helices (αA:Pro346-Ser350; αB: His372-Asn381) and six β-strands (βA to βF) arranged in a β-barrel (Figure 1, left). The same structure in complex with a target peptide is shown in Figure 1, right. The peptide sits in a groove between the αB helix and the βB strand as in the typical binding geometry common to all PDZ domains.

The general features of PDZ domains are the presence of a Gly-X-Gly-X motif in the loop capping the binding groove (loop L1). For PDZ3 the two X residues are a Leu-323 (L) and Phe-325 (F), which are at the origin of the widespread GLGF name of PDZ domains (the residue numeration is according to the 1bfe Protein Data Bank file(Berman and others 2000)). The loop L1 also contains a basic residue (Arg-318 in PDZ3, which in other sequences can be substituted with a lysine) that forms a water mediated hydrogen-bond with the carboxylate terminal of the bound peptide. A hydrophobic pocket that specifically targets peptides with hydrophobic C-terminal amino-acids is lined by residues Leu-323, Phe-325, Ile-327 and Leu-379 that can be different in other PDZ domains, although they are invariably hydrophobic. The high specificity for different targets is instead provided by the higher variability of the other amino-acids forming the αB helix and the βB strand (Bezprozvanny and Maximov 2002; Doyle and others 1996).

Peptide binding induces a structure deformation involving a concurrent displacement of the loop L1, of the αB helix and, to a lesser extent, of other secondary motifs. An effective representation of the correlations among these structural rearrangements can be obtained through the cross-correlation matrix between Cα carbon atoms

$$C_{ij} = \sum_{\alpha=x,y,z} d_{i\alpha} \cdot d_{j\alpha} \qquad (8)$$

where $\vec{d}$ is the normalized vector of binding induced atomic displacements (see Methods), and $d_{i\alpha}$ represents the $\alpha$-th component of the rescaled difference between the positions of the $i$-th Cα atom in the complexed and peptide-free PDZ3 structures (see Eq.4). The color-code representation of $C_{ij}$ is shown in Figure 2. A red (blue) entry of $C_{ij}$ indicates (anti)parallel distortions of the $i$-th and $j$-th Cα atoms.

The analysis of the $C_{ij}$-matrix reveals that the L1 loop performs an anti-parallel motion with respect to the αB helix and with large portions of the overall PDZ3 structure. The entries along the diagonal maintain the information about deformation amplitudes thanks to the overall normalization (Eq.4). They can be used, therefore, to highlight the L1 loop and the αB helix as the regions undergoing the largest deformations upon binding. These results hint at a binding mechanism that opens the hydrophobic pocket, which is located between the αB helix and the βB strand and covered by the L1 loop.

***Normal Modes Analysis***

The normal modes (NM) of a protein structure describe the collective displacements of its atoms around their equilibrium positions (local energy minimum). Each NM is characterized by a frequency of vibration that determines the characteristic relaxation time in the case of overdamped dynamics, and any small deformation of the original structure can be exactly described as a linear superposition of its NMs. Here we apply NMA to describe the collective deformations of various PDZ domains around their equilibrium structures as obtained from X-ray crystallography or solution NMR.

## A) CHARMM Force-Field

We performed the NMA of the PDZ3 domain by means of the corresponding module of the CHARMM package (see Methods). A close examination of the CHARMM results requires a detailed inspection of the eigenvectors (the Normal Modes) of the Hessian matrix of the CHARMM force field computed at the relaxed native structure (see Methods). To characterize how different NMs are involved in the binding deformation, we computed both their *involvement coefficients* $I_n$ and their *thermal involvement coefficients* $T_n$ (see Methods) that describe the contribution of each mode to the overlap of the binding deformation with a typical fluctuation of the structure. We found that the binding deformation has maximal thermodynamic overlap with the second and third non-trivial CHARMM NMs (notice that the first six roto-translational eigenvectors of the Hessian matrix have been discarded), hereafter referred to as M2 and M3 respectively. Figure 3 shows the $T_n$'s, normalized in such a way that $\sum_{n=1}^{3N-6} T_n^2 = 1$. The second and third NMs account together for roughly 47% of the overlap of the binding distortion with the thermal fluctuations of the structure. This means that the distortion is unevenly distributed among the different modes; rather it privileges two low-frequency ones. Furthermore, although the $T_n$'s intrinsically reward low frequency modes, the exceptionally high contributions of the second and third modes, and modest participation of the first and fourth, indicate that indeed the geometry of the modes plays a very important role.

Correlations of the motions of Cα's within modes M2 and M3 are represented by a color-coded matrix in Figure 4. M2 involves the concerted motion of the αA helix and of the L1 loop, with minor movements of other secondary structures. It does not produce, therefore, a significant rearrangement of the binding region, although the large L1 motion can clearly cap/uncap the binding groove. Mode M3 shows a pattern very similar to the binding deformation (Figure 2) and it displays remarkable movements of the L1 and L6 loops and of the αB helix. The group αB+L6 moves in an antiparallel way with respect to L1 as can be seen by the boxed blue region in Figure 4. Specifically, M3 describes the opening/closing of the hydrophobic pocket.

The high correlation of such two modes with the binding deformation, and their simple and intuitive interpretation, suggests that they assist and facilitate the binding. Thus, the presence of a few modes that best couple to the deformation implies that, as the binding proceeds, the whole binding groove opens and closes *collectively* in the correct way.

## B) Cβ-Anisotropic Network Model: focus on PDZ topology

The highly conserved structure of PDZ domains strongly points to a topology well suited for the binding process. To explore the different role of the PDZ topology with respect to its chemistry in determining the binding movements, we have shifted from the CHARMM force-field, where both topology and full chemical details are taken into account, to the ANM description in its C$\beta$ version, which deals only with the topological features of the domain (see Methods).

We then repeated the analysis of the NMs by computing the geometric and thermodynamic overlap between the binding deformation and the NMs. Results are summarized in Figure 5, where again the second and third NMs emerge as the best coupled to the PDZ function. Together, they account for roughly 65% of the whole binding deformation. The internal correlations of the second NM (Figure 6) are remarkably similar to the binding deformation (Figure 2) and again select an antiparallel motion of the L1 loop and of the αB+L6 group, leading to the "breathing" of the binding groove, just as the CHARMM M3 mode.

The similarity of CHARMM and C$\beta$-ANM analysis and their agreement with the deformation induced by the binding suggest a prominent role of the topology of the domain in determining its functional dynamical behavior.

### Dynamic information from experimental data

The agreement between the CHARMM and C$\beta$-ANM NMA extends to the computation of the B-factors, as it can be seen from lower panel of Figure 7. Instead, the B-factors from crystallographic experiments (1bfe) (Doyle and others 1996) are significantly different from the theoretical ones (Figure 7, upper panel). We argue that such a discrepancy rests on the possible effects due the crystal. To check this hypothesis we have studied the PDZ domain of the ZASP protein, whose structure has been recently solved by NMR (Au and others 2004) (see Methods). ZASP-PDZ is a suitable candidate for this check since its reconstruction has been modeled on the PDZ3 template (1bfe). Comparison between theoretical and experimental Root Mean Square Displacements (*rmsd*) is shown in Figure 8. The atomic rmsd's from NMR are due to experimental uncertainties on the restraints used for the structure refinement, and have various origins among which the intrinsic dynamic flexibility of the protein. In Figure 8, we show that the C$\beta$-ANM B-factors are close to the rmsd of the ZASP-PDZ structure, and strongly suggest on the one hand that the rmsd's are likely to be dynamical in nature, and on the other hand that the dynamical information from the X-ray data could indeed be distorted by the crystal context. The relevance of the crystal has been explored previously (Eastman and others 1999) showing that vibrational B-factors can change significantly between proteins in solution and in crystals, where, moreover, the total B-factors can be affected by several other contributions such as roto-translations of the whole molecule in the lattice, conformational substates and crystal disorder.

### Collective dynamics of the PDZ domain

Recent NMR experiments on the second PDZ domain (PDZ2 hereafter) of human protein tyrosine phosphatase 1E (hPTP1E) have pointed out that the binding process influences the dynamics of residues that are not in direct contact with the target peptide (Fuentes and others 2004). In particular experiments have identified two regions of the PDZ domain, comprising eight amino-acids (out of a total 85), whose dynamics is most influenced by the binding. Since there seems to be no large conformational changes associated to the binding process, NMA is an appropriate tool to probe the correlation between the collective dynamics of PDZ2 and the binding process. We have computed the Cβ-ANM NMs of the peptide-free structure, and their overlap with the binding induced deformation (Methods). In this case (Figure 9) the NM with the highest involvement coefficient (both purely geometrical and thermodynamic) is the 9-th one. All of the eight amino-acids identified in (Fuentes and others 2004) fall close to some of the most mobile regions of this mode (Figure 10). The mobile regions correspond to the six peaks in Figure 10 where the squared amino-acid displacements are above their average (dotted line), and we extend them to take into account two positions both on the left and on the right of each peak to take into account also the hinges. The probability that seven amino-acids chosen at random over the whole PDZ2 sequence <u>all</u> belong to these regions is roughly 0.08. Anyway also notice that Ala-79, which does not belong to the mobile regions, is still in their proximity. Such a small probability clearly indicates that the nice matching of the mobile regions with the affected amino-acids from Fuentes *et al.* ((Fuentes and others 2004)) is unlikely to occur by chance. Rather, it signals that the 9-th NM, chosen according to structural information, also captures the global binding dynamics of the domain, and as such, the matching is not surprising at all, since there is a strong dynamical correlation between the observed NMR data and the selected NM.

## Conclusions

Conformational changes, with or without major structural rearrangements, are fundamental to protein functions and are often associated to binding of ligands such as ions, nucleotides (ATP and GTP) and peptides. In recent years, a dynamical picture of these transformations has emerged, showing that each different conformation is usually populated at equilibrium, according to its free energy. The role of ligands is to shift, in a not exclusive way, the populations toward one specific conformation (for a recent review see Kern and Zuiderweg (Kern and Zuiderweg 2003)). Consequently, the dynamics acquires a much more important role implying that proteins, fluctuating around their equilibrium states, are able to pre-explore most of the different functional states. NMA of various PDZ domains shows that the deformation associated with the binding of a peptide is nicely captured by a single or at most a couple of low-frequency collective domain movements. The protein, therefore, dynamically explores those different conformations relevant to the function in an optimally coordinated way, since the movements of individual residues are strictly correlated within each NM. In other words, through thermal fluctuations PDZ domains are therefore ready for the binding. Even when the binding/unbinding process would be influenced by anharmonic movements, we are confident that Normal Modes represent non-trivial precursors.

Even though NMR experiments often need the X-ray structure as a template, according to our analysis they compare better with our theoretical predictions and this could be perhaps a hint that NMR provides more reliable information on the dynamics of proteins.

Furthermore our NMA helped to rationalize the recent findings (Fuentes and others 2004) that peptide binding influences the dynamics of residues that are not in direct contact with the ligand. Our results confirm that peptide binding affects the whole domain structure, because it is strongly coupled to a single normal mode which is intrinsically collective, in agreement with previous speculations (Cooper and Dryden 1984).

# Figure captions

**Figure 1.** Cartoon representation of the X-ray structure of the PDZ3 domain of PSD-95 without (left) and with (right) bound peptide. Secondary structures are highlighted by different colors: red for α-helices, green for β-strands and blue for loops. The bound peptide (orange, right figure) is shown in stick representation, apart from the C-terminal valine, all the atoms of whom are shown to emphasize that it sits in a hydrophobic pocket lined by Leu-323, Phe-325, Ile-327 and Leu-379 (pink all-atom representation, right figure).

**Figure 2.** Internal Cα-Cα correlations of the distortion induced by the binding for the PDZ3 domain of PSD-95, computed according to equation 1. Red (blue) regions correspond to (anti)parallel distortions. The loop L1 and helix αB are the most mobile regions of the protein, and they are mutually antiparallel. The loop L1 displacement is actually antiparallel to most other elements of the protein. Note that due to the overall normalization of the displacement vector, the diagonal elements are not unitary but their sum.

**Figure 3.** The thermal involvement coefficients $T_n$ of the Charmm modes of the PDZ3 domain of PSD-95. The squares of the displayed values represent their percentage contributions to the thermodynamic fluctuations of the overlap between the binding deformation and a thermodynamic structure fluctuations (equation 8). Inset: the usual involvement coefficients $I_n$ (equation 7).

**Figure 4.** Internal Cα-Cα correlations for M2 (left) and M3 (right) Charmm vibrational normal modes of the PDZ3 domain of PSD-95. M3 (right) shows a pattern rather similar to the internal correlations of the binding distortion (Figure 2).

**Figure 5.** The thermal involvement coefficients $T_n$ of the Cβ-ANM modes of the PDZ3 domain of PSD-95. The squares of the displayed values represent their percentage contributions to the thermodynamic fluctuation of the overlap between the binding deformation and a thermodynamic structure fluctuation (equation 8). Inset: the usual involvement coefficients $I_n$ (equation 7).

**Figure 6.** (Left) Internal Cα-Cα correlations for the second Cβ-ANM normal mode of the PDZ3 domain of PSD-95. They are remarkably similar to the internal correlations of the binding deformation. (Right) The regions most affected by the displacements are represented in blue (loop L1) and red (helix αB), with the peptide shown in orange.

**Figure 7.** Comparison of experimental (X-ray) and theoretical (Charmm and Cβ-ANM) B-factors of the PDZ3 domain. Whereas the Charmm and Cβ-ANM B-factors show a very remarkable mutual agreement (lower panel), they are much less in agreement with the X-ray ones. B-factors are expressed in $Å^2$.

**Figure 8.** Rmsd of the solution structure of the Cα atoms of the ZASP PDZ domain (solid line) compared with the theoretical ones computed with the Cβ-ANM model (dashed line). The match is not the result of a fit, which would be extremely skewed by the regions around residues 10 and 70.

**Figure 9.** The thermal involvement coefficients $T_n$ of the Cβ-ANM modes of the PDZ2 domain of hPTP1E. The squares of the displayed values represent their percentage contributions to the thermodynamic fluctuation of the overlap between the binding deformation and a thermodynamic structure fluctuation (equation 8). Inset: the usual involvement coefficients $I_n$ (equation 7).

**Figure 10.** Square fluctuation amplitudes of the ninth Cβ-ANM normal mode of the PDZ2 domain of hPTP1E. The dotted line is the average square displacement. The circles mark the amino-acids that are dynamically most affected by the binding process, and they are labeled by their amino-acid species (1-letter code for clarity).

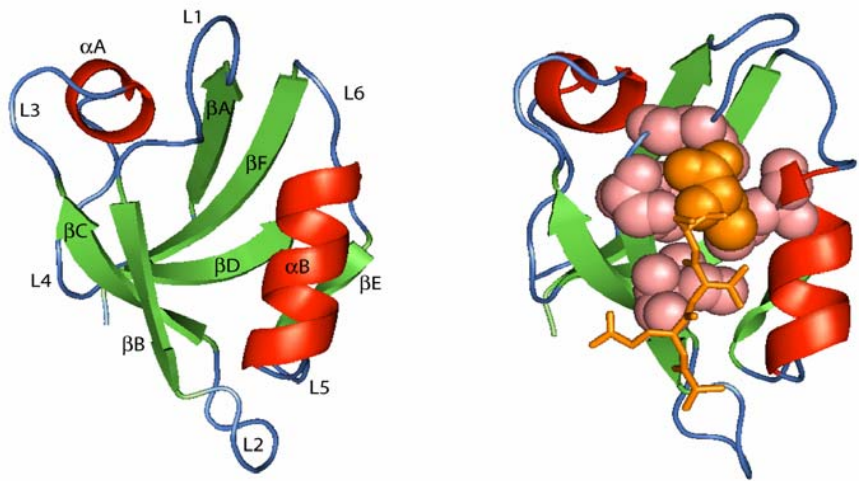

**P. De Los Rios et al.     Figure 1**

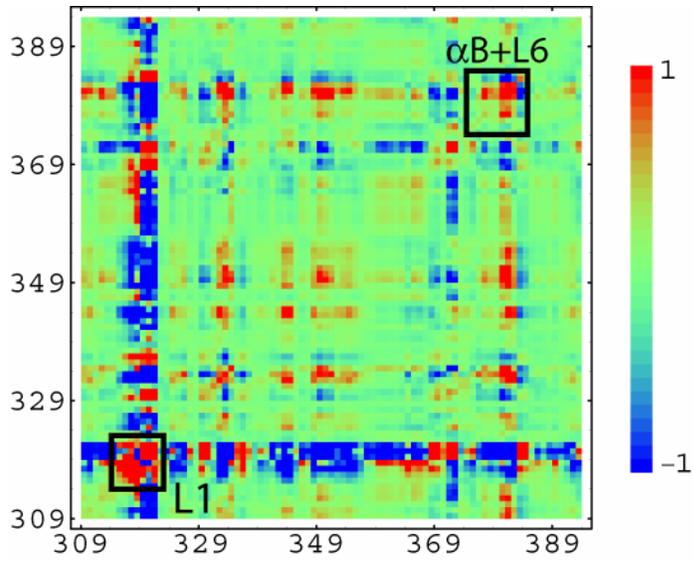

P. De Los Rios et al.    Figure 2.

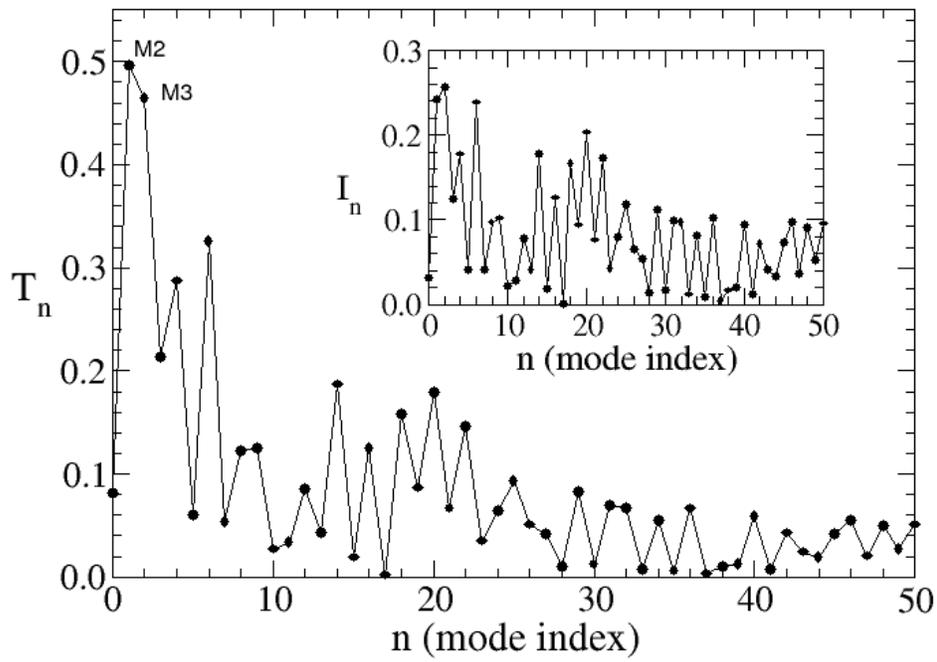

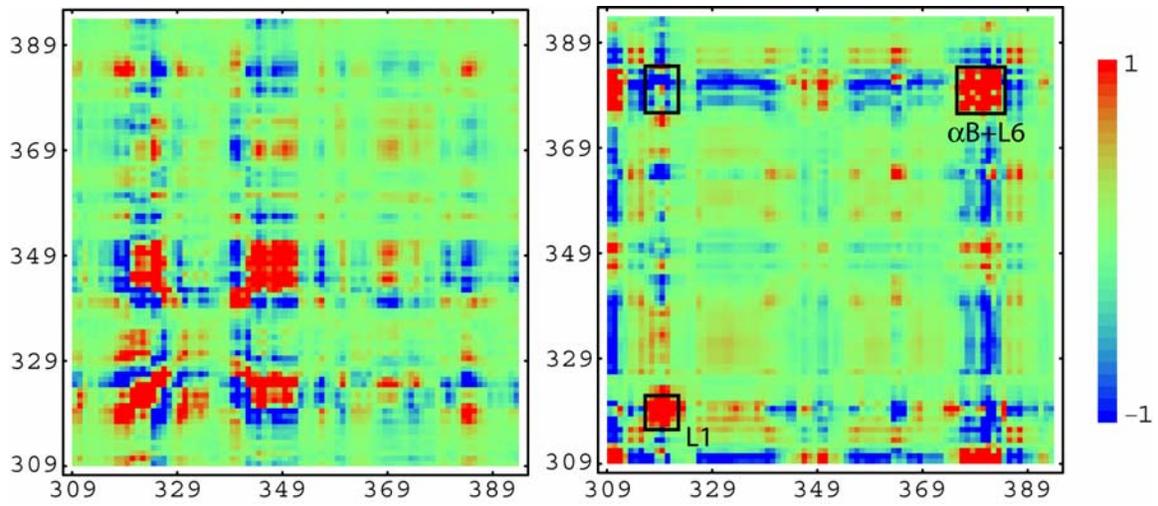

**P. De Los Rios et al.    Figure 4.**

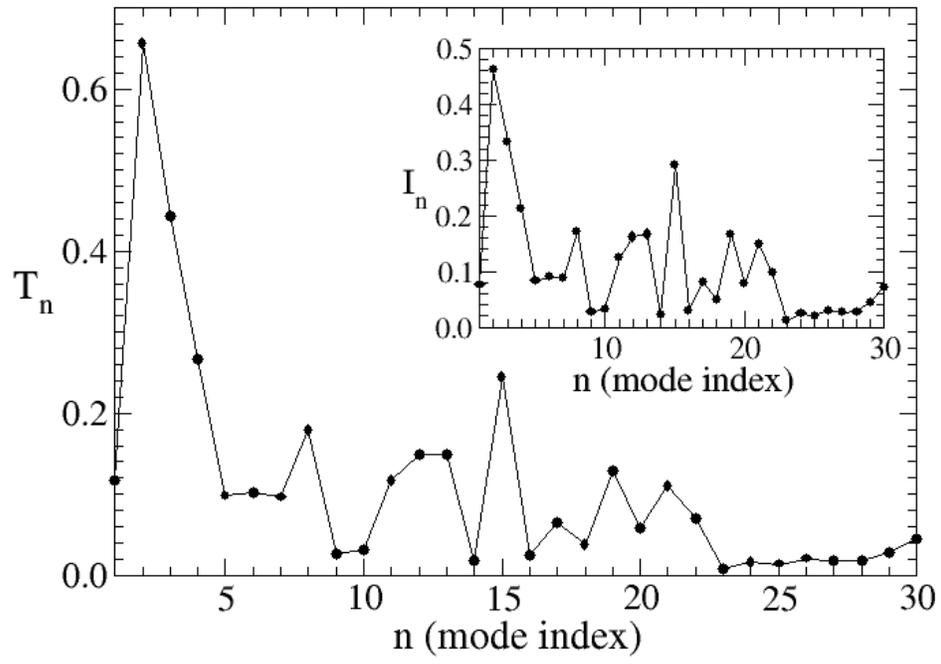

P. De Los Rios et al.    Figure 5.

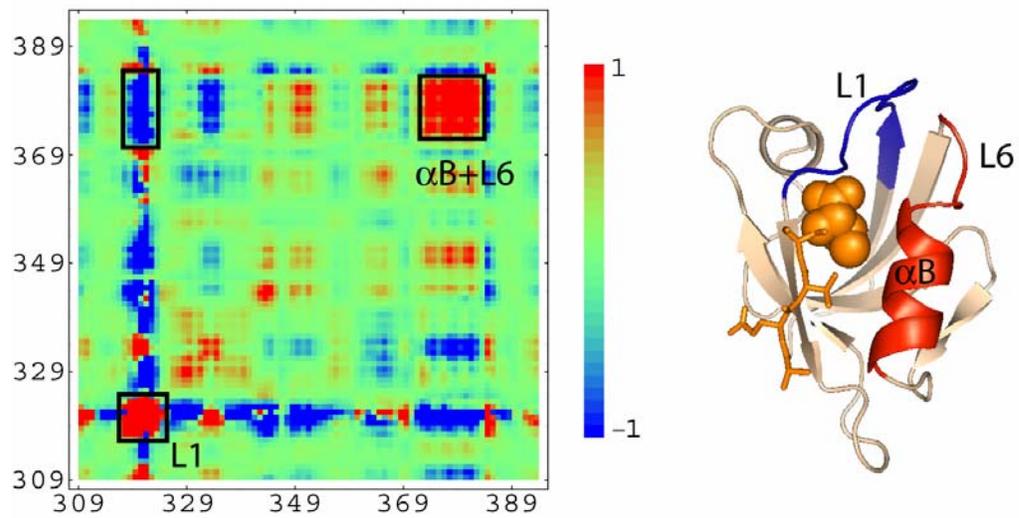

**P. De Los Rios et al.     Figure 6.**

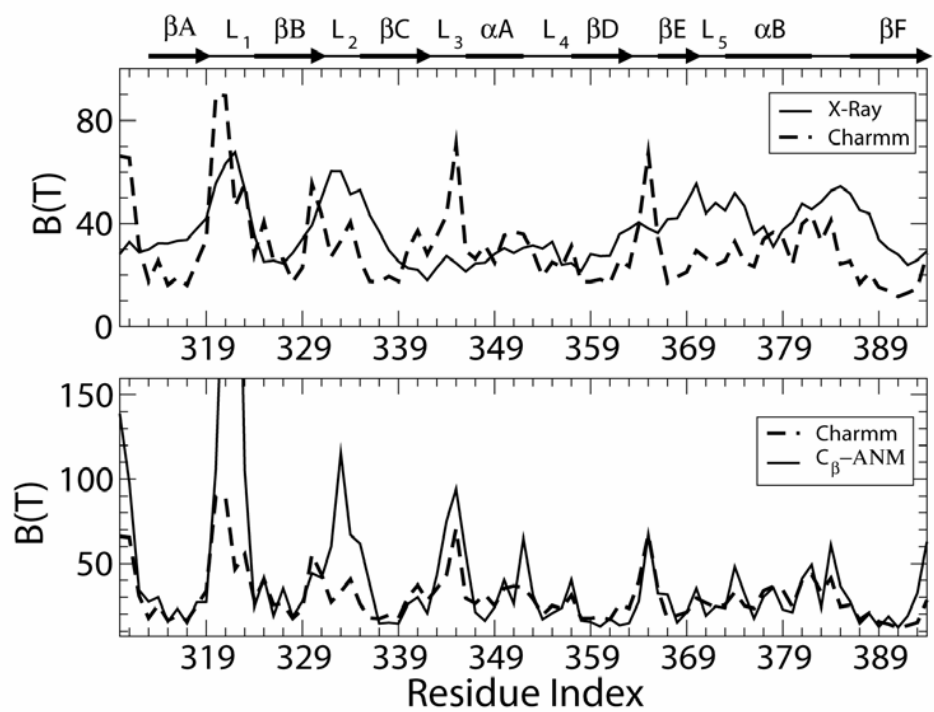

P. De Los Rios et al.    Figure 7.

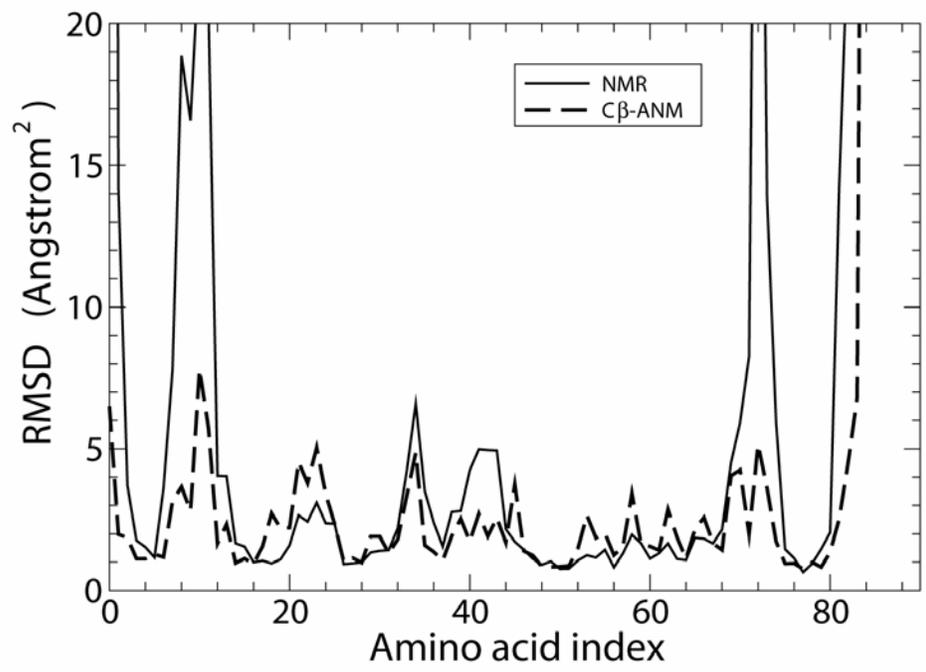

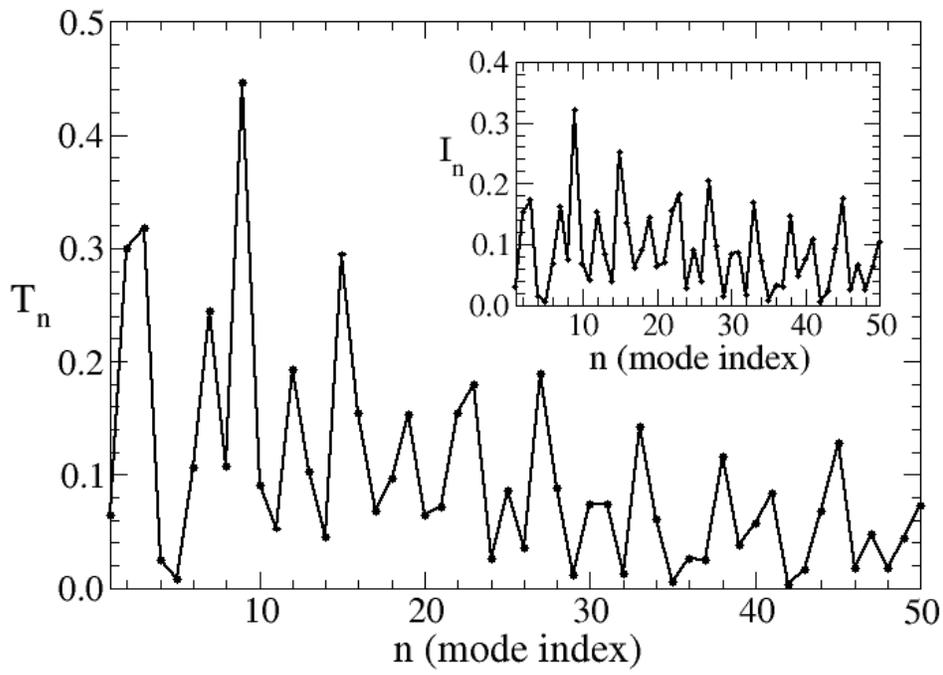

P. De Los Rios et al.     Figure 9.

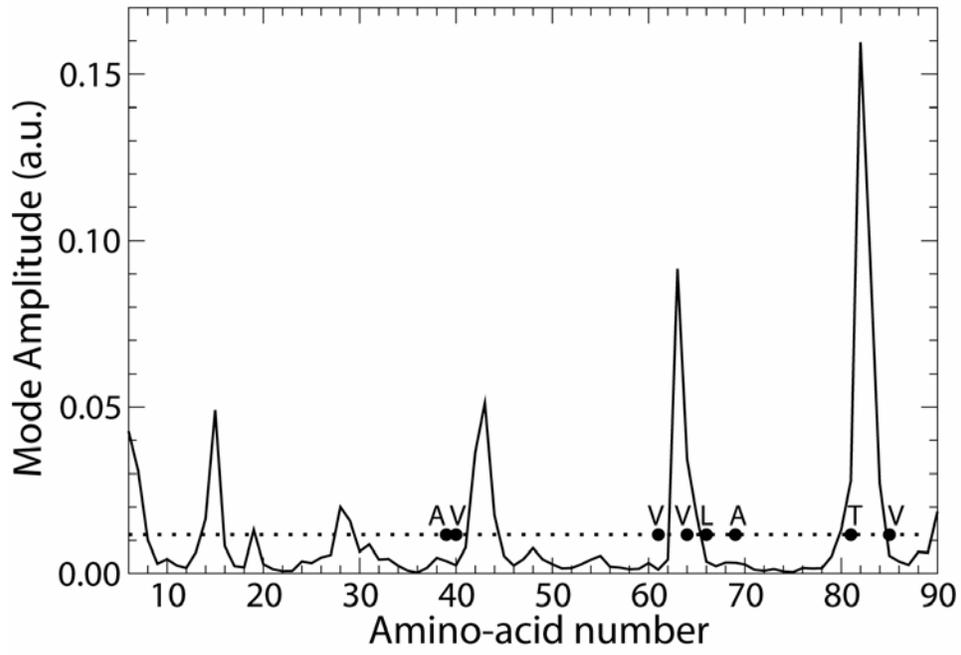

P. De Los Rios et al.        Figure 10.